\documentclass[fleqn,usenatbib]{mnras}

\usepackage{newtxtext,newtxmath}
\usepackage[T1]{fontenc}

\DeclareRobustCommand{\VAN}[3]{#2}
\let\VANthebibliography\thebibliography
\def\thebibliography{\DeclareRobustCommand{\VAN}[3]{##3}\VANthebibliography}

\usepackage{graphicx}	
\usepackage{amsmath}	
\usepackage{soul}
\usepackage{bm}

\title[Identifying lensed GWs with 3G detectors]{Identifying Strongly Lensed Gravitational Waves with the Third-generation Detectors}

\author[Gao et al.]{
Zijun Gao,$^{1}$
Kai Liao,$^{1}$\thanks{E-mail:liaokai@whu.edu.cn}
Lilan Yang$^{2}$
and Zong-Hong Zhu$^{1,3}$\thanks{E-mail:zhuzh@whu.edu.cn}
\\
$^{1}$Department of Astronomy, School of Physics and Technology, Wuhan University, Wuhan 430072, China\\
$^{2}$Kavli Institute for the Physics and Mathematics of the Universe (Kavli IPMU, WPI), University of Tokyo, Chiba 277-8583, Japan\\
$^{3}$Department of Astronomy, Beijing Normal University, Beijing, 100875, China
}

\date{Accepted XXX. Received YYY; in original form ZZZ}

\pubyear{2023}

\begin{document}
\label{firstpage}
\pagerange{\pageref{firstpage}--\pageref{lastpage}}
\maketitle

\begin{abstract}
The joint detection of GW signals by a network of instruments will increase the detecting ability of faint and far GW signals with higher signal-to-noise ratios (SNRs), which could improve the ability of detecting the lensed GWs as well, especially for the 3rd generation detectors, e.g. Einstein Telescope (ET) and Cosmic Explorer (CE).  
However, identifying Strongly Lensed Gravitational Waves (SLGWs) is still challenging.
We focus on the identification ability of 3G detectors in this article. 
We predict and analyze the SNR distribution of SLGW signals and prove only 50.6\% of SLGW pairs detected by ET alone can be identified by Lens Bayes factor (LBF), which is a popular method at present to identify SLGWs.
For SLGW pairs detected by CE\&ET network, owing to the superior spatial resolution, this number rises to 87.3\%. Moreover, we get an approximate analytical relation between SNR and LBF. 
We give clear SNR limits to identify SLGWs and estimate the expected yearly detection rates of galaxy-scale lensed GWs that can get identified with 3G detector network. 

\end{abstract}

\def\mc{\mathcal}
\def\Pr{{P}}
\def\vtheta{{\bm \theta}}
\def\Zl{{\mc{Z}_\textsc{l}}}
\def\Zu{{\mc{Z}_\textsc{u}}}
\def\Blu{{\mc{B}_\textsc{u}^\textsc{l}}}



\section{Introduction}

There is an interesting similarity between gravitational waves (GWs) and electromagnetic waves.
They have the same wave equation in weak field approximation and follow the same propagation path i.e. null geodesic line, which provides the theoretical foundation for the optical gravitational lensing as well as the gravitational lensing theory of GWs. 

Strongly lensed gravitational waves (SLGWs), due to the gravitation of galaxies and clusters between sources and detectors \citep{2015JCAP...12..006D,2018MNRAS.475.3823S,PhysRevD.97.023012}, are very ideal objects to develop cosmological studies. 
Since the first GW detection from a binary black hole merger~\citep{PhysRevLett.116.061102}, a tide of searching and researching for lensed gravitational wave signals was followed. For instance, the detection of lensed GW events enables investigation of the properties of the binaries\citep{con, 2021arXiv211103634T}, breaking the mass-sheet degeneracy in lensing systems \citep{2021PhRvD.104b3503C}, figuring out the binary formation channel \citep{10.1093/mnras/stab1980}, testing General Relativity~\citep{2017PhRvL.118i1101C,2017PhRvL.118i1102F,Goyal_2021,hernandez2022measuring}, constraining cosmological parameters~\citep{2017NatCo...8.1148L,Hannuksela_2020}, probing compact dark matter/primordial black holes~\citep{2019PhRvL.122d1103J,2020MNRAS.495.2002L} and so on. 
Many works have been done to search for SLGWs. However, there has yet to be any compelling proof to support their existence.\citep[e.g.][]{2020arXiv200712709D,2022arXiv220608234K,2019ApJ...874L...2H,Abbott_2021,2023arXiv230408393T,Liu_2021}.
For more details on lensed GWs, we refer to the reviews~\citep{2019RPPh...82l6901O,2022ChPhL..39k9801L}.\par

The detection of SLGWs is limited by poor detectability of current detector network. 
On one hand, the current detection is based the 2nd generation GW detector network which consist of 
LIGO Handford, LIGO Livingston, Advanced Virgo (HLV) and later joined KAGRA. 
The third observation run (O3) has accomplished and only tens of detections have been reported, 
the event rate is still relatively low.
Even worse, the probability of SL is low, approximatly $\sim$0.1\% \citep{10.1093/mnras/stab3298},
which results in no compelling proof of SLGWs detection.  
On the other hand, the low signal-to-noise ratio (SNR) causes a low confidence level of the GW parameter estimation, which is important  
to identify strongly lensed gravitational waves \citet{2021arXiv210409339L,2018arXiv180707062H}. 
The parameter estimation with a low confidence level will make it hard to determine whether a set of signals are lensed or unlensed. 
It is challenging to determine whether a set of signals are lensed or unlensed based on poor GW parameter constraints. We will present more details in section \ref{sec:SNR and LBF}.

Fortunately, the detect-ability will be significantly improved 
by third-generation (3G) detectors, i.e., the Einstein Telescope (ET) and Cosmic Explorer (CE).
Their improved sensitivity to lower frequencies will lead to better detection rates and higher SNRs of signals. 
These will help the studies of lensed gravitational waves in many aspects. 
Event rates of strongly lensed gravitational waves will increase to hundreds per year \citep{10.1093/mnras/stab3298}. 
Different types of images will be distinguishable \citep{2021PhRvD.103j4055W}. But the identification of SLGWs is still challenging.

In this paper, 
we aim to predict the identification of SLGWs based on the 3G detector network and search for the relation between SNR and the ability to identify SLGWs.
We utilize Monte Carlo simulation to generate thousands of detectable strongly lensed GW signal sets and calculate the SNRs of a part of sets, and this method is based on the previous work \citet{10.1093/mnras/stab3298}.
Most of detected GW events are from binary black hole (BBH) mergers. Therefore, we only discuss potential SLGWs from BBHs in this paper.
To distinguish the lensed and unrelated events from BBH merger events,
We use the method in \citet{2018arXiv180707062H} to calculate Lens Bayes factors: the ratio of the marginal likelihoods of GW signal sets under the lensed and unlensed hypotheses.
Our goal is to provide clear SNR criteria to restrict the expected yearly detection rate of galaxy-scale lensed GWs that can get identification.

This paper is organized as follows. We introduce the methodology of generating simulated SLGW signals, detecting and parameter estimating a set of SLGW signals emitted from binary black holes (BBHs) in Section 2. The Bayesian statistical framework is discussed in Section 3. Results and discussions are presented in Section 4. Conclusions and outlook are given in Section 5.

\section{Simulation of GW signals}
This section introduces generating SLGW signals by galaxies via Monte Carlo, the detection and parameter estimation.    

\subsection{Generation of SLGW signals } 
One of our goals is to forecast the event rate of SLGWs based on 3G detector network. We mock the detection scenario as realistic as possible via Monte Carlo method. The details have been given in previous work \citet{10.1093/mnras/stab3298}, and here we briefly summarize the method. We first introduce how to simulate a GW signal in Section~\ref{sec:mock_gw}, then take gravitational lensing effect into consideration, see Section~\ref{sec:lensing}.

\subsubsection{Mock GW signals}\label{sec:mock_gw}
First, we simulate the signal of GW based on templates. 
Quasi-circular orbit CBC waveform depend on 15 parameters: two component masses: $m_1$, $m_2$; six component spins: $\vec{s_1}$, $\vec{s_2}$; 2-D sky location: right ascension $\mathrm{RA}$, declination $\mathrm{DEC}$, inclination angle $\theta_{JN}$, polarization angle $\psi$, luminosity distance $d_{\mathrm{L}}$, and coalescence phase $\phi_c$, time $t$. In our research, we only consider situations in which orbital angular momentum and spin angular momentum are collinear. So we only need 11 parameters without parameters of direction angle of spin angular momentum $\theta_1$, $\theta_2$, $\Delta\phi$, and $\phi_{JL}$. Table \ref{tab:bbh} shows their distributions for generating GW signals. These distributions are similar to the TABLE I. in \cite{PhysRevLett.125.101102}.  

We use the intrinsic merge rates for the BBH systems as a function of cosmological Redshift according to Dominik et al.\citet{Dominik_2013}.  The proper merging rate is:
\begin{equation}
 \frac{d\dot{N}}{{dz_{\mathrm{s}}}{d\Omega }}
 =\frac{d_H}{E(z_{\mathrm{s}})}{{r}^2(z_{\mathrm{s}})}\frac{\dot{n}_{0}(z_{\mathrm{s}})}{1+z_{\mathrm{s}}},
 \end{equation}
where $E(z_{\mathrm{s}})\equiv\frac{H(z_{\mathrm{s}})}{H_0}$ is the dimensionless Hubble parameter at $z=z_{\mathrm{s}}$. $\dot{n}_{0}(z_{\mathrm{s}})$ is intrinsic merge rates at $z_{\mathrm{s}}$ per comoving space. In \citet{Dominik_2013}, they adopted the formula provided by \citet{Strolger_2004} to get the star formation rate. They employed two distinct scenarios for metallicity evolution, with median value of metallicity of 1.5 $Z_{\sun}$ and 0.8 $Z_{\sun}$ at $z\sim0$ labeled as high-end and low-end, respectively. They also provided a suite of evolutionary models including 
the standard one and three of its modifications, i.e., Optimistic Common Envelope, delayed SN explosion and high BH kicks scenario to get the merge rates. In this paper, we only consider the low-end metallicity, standard scenario as evolution model as it has been considered in all previous works.

$d_H$ is the Hubble distance defined as $d_H=\frac{c}{H_0}$.${r}(z_{\mathrm{s}})$ is the comoving distance defined as ${r}(z_{\mathrm{s}}) = d_H \int_0^z \frac{dz'}{E(z')}$. $z_{\mathrm{s}}$ is the source redshift and $\Omega$ is solid angle. We finally adopted the redshift distribution as \cite{2020MNRAS.495.2002L}, calculated by the StarTrack evolutionary code. \footnote{We used the data available at the website \url{http://www.syntheticuniverse.org}.}.  

We use flat $\Lambda$-CDM Model with $H_0 = \rm{70 \;km\;s^{-1}\;Mpc^{-1}}$ and $\Omega_{\rm{M}}=0.307$, and we generate luminosity distances of sources from redshift distribution using: $d_{\mathrm{L}}^{\mathrm{s}}=r(z_{\mathrm{s}})(1+z_{\mathrm{s}})$. Considering the influence of lensing, a given image has an absolute magnification $\mu $ which can be defined in terms of the apparent luminosity distance $d_\mathrm{L}$ as $d_\mathrm{L}=\frac{d_\mathrm{L}^{\mathrm{s}}}{\sqrt{\mu}}$.

\subsubsection{Gravitational lensing effects}\label{sec:lensing}
Due to the gravitation of galaxies and clusters between GW signal and detectors, part of the GW events will be lensed. 
Here we consider the simplest lens model, i.e., singular isothermal sphere (SIS), to discuss the lensing effects. 
The optical depth of the SIS lenses can be calculated as:

\begin{equation} \label{eq:tau}
\small
\tau_0(z_{\mathrm{s}}) = \frac{1}{4 \pi} \int_0^{z_{\mathrm{s}}} dz_\mathrm{l}  \int^{\infty}_0 d \sigma\;
4 \pi \left( \frac{c}{H_0} \right)^3 \frac{ {\tilde r}_\mathrm{l}^2}{E(z_\mathrm{l})} S_{\mathrm{cr}} \frac{d n}{d \sigma},
\end{equation} 
where $S_{\mathrm{cr}}=16 \pi^3 \left( \frac{\sigma}{c} \right)^4 \left( \frac{{\tilde r}_\mathrm{ls}} {{\tilde r}_\mathrm{s}} \right)^2$ is the scattering cross section of gravitational lensing with SIS model. In this paper we take the velocity dispersion distribution function $\frac{d n}{d \sigma}$ from \citet{2007ApJ...658..884C}. 
In the context of SIS, there are two lensed images with different magnifications, hence different SNRs. The one with higher SNR will be created at angular position $\theta_+$. It is outside the Einstein radius ($|\theta_+| > \theta_\text{E}$), whose magnification is $\mu_+=\sqrt{\left|\frac{\theta_+/\theta_\text{E}}{|\theta_+/\theta_\text{E}|-1}\right|}$.The one  with lower SNR will be created at angular position $\theta_-$. It is inside the Einstein radius ($|\theta_-| < \theta_\text{E}$), whose magnification is $\mu_-=\sqrt{\left|\frac{\theta_-/\theta_\text{E}}{|\theta_-/\theta_\text{E}|-1}\right|}$. We define $\beta$ as the angular position of the source. With SIS model we have $\theta_\pm=\beta\pm\theta_\text{E}$. Note that the angular position of the source must be inside the Einstein radius ($\beta<\theta_\text{E}$).   Furthermore, there will be a time delay $\Delta t$ between two signals due to the deflected GWs.  
The `$\theta_+$'arrives first and the `$\theta_-$' follows. 
Due to the rotation of the earth, two signals with different arrival times will be modulated by detector antenna response functions (\ref{eq:antenna}) in diurnal and annual periodic \citep{Yang_2019}, so we generate $\Delta t$ randomly within $30 \times 86400s$.

\begin{table}
\centering
\caption{The distributions of parameters for generating GW siganls. $m_1^{\mathrm{s}}$ and $m_2^{\mathrm{s}}$ is the source masses of each BH in source coordinate system. $a_1$, $a_2$ are dimensionless spin of each BH defined as $a=\frac{cJ}{GM^2}$, RA is the right ascension, DEC is the declination, $\theta_{JN}$ is the orbital inclination, $\psi$ is the polarization angle, $\phi$ is the phase, $\delta t$ is the difference of geocent time, and $\frac{\beta}{E_\theta}$ is the dimensionless angular position of source in SIS lensing model.}

\begin{tabular}{lccccc}
\hline
\hline
parameters & unit & shape & min & max \\
\hline
$m_1^{\mathrm{s}}$   & $M_{\sun}$ & Power-law, $\alpha=-2.3$ & 5 &  50 \\
$m_2^{\mathrm{s}}$   & $M_{\sun}$ & Power-law, $\alpha=-2.3$ & 5 &  50 \\

$a_{1}$ (BBH) & 1 & Uniform & -0.99 & 0.99 \\
$a_{2}$ (BBH) & 1 & Uniform & -0.99 & 0.99 \\
RA &  rad. & Uniform &   0    &  $2\pi$    \\
DEC & rad. & Cos & $-\pi/2$  & $\pi/2$ \\
$\theta_{JN}$ & rad. & Sin & 0 & $\pi$ \\
$\psi$ & rad. & Uniform & 0 & $\pi$ \\
$\phi_c$ &  rad. & Uniform & 0 & $2\pi$\\
$\delta t$ & s & Uniform & 0 & $3600\times24\times30$ \\
$\frac{\beta}{E_\theta}$ & 1 & Uniform & 0 & 1 \\
\hline
\label{tab:bbh}
\end{tabular}
\end{table}

\subsection{GW signal detection and parameter estimation}
\label{sec:detection and parameter estimation of GW signal}
We have introduced how to mock a SLGW signal, and in the section, we will discuss the detection and parameter estimation of the signal.

The common method for identifying a GW signal from CBC is template-based searches. The matched filtering method is the key to template-based searches. We need to define the matched-filter signal to noise ratio (SNR):
\begin{equation}\label{eq:snr}
\rho^2(t) = \frac{1}{C}|\frac{1}{4}\int_{f_\mathrm{min}}^{f_\mathrm{max}}\frac{\tilde{s}(f)\tilde{h}(f)^\ast}{S_n(f)}e^{2\pi{i}ft}df|^2,
\end{equation}
where \begin{equation}\label{eq:nom}
C = 4\int_{f_\mathrm{min}}^{f_\mathrm{max}}\frac{\tilde{h}(f)\tilde{h}(f)^\ast}{S_n(f)}df,
\end{equation}
and the $\tilde{h}(f)$ is the template's strain in the frequency-domain. $\tilde{s}(f)$ is the signal in frequency-domain (i.e. Fourier transformed signal in time-domain ). $S_n(f)$ is the one-sided power spectral density of the GW detector. $f_\mathrm{min}$ is the minimum-frequency or in other words, a cut off frequency. We choose 20 Hz in this study. 

If a signal satisfies $\rho > \rho_0$, where $\rho_0$ is a identification threshold, we then identify the signal as a GW signal. We choose the threshold as $\rho_0 =8$ in this paper.

Detectors respond to GW signals in terms of antenna pattern function:
\begin{equation}\label{eq:antenna}
h_{\mathrm{det}}=F_{+}(\theta,\phi,\psi)h_{+}+F_{\times}(\theta,\phi,\psi)h_{\times}
\end{equation}
where
\begin{equation}\label{eq:antenna_sub1}
F_{+} = \frac{1}{2} (1 + \cos^2{\theta}) \cos{2\phi} \cos{2 \psi} - \cos{\theta} \sin{2 \phi} \sin{ 2 \psi}
\end{equation}
\begin{equation}\label{eq:antenna_sub2}
F_{\times} = \frac{1}{2} (1 + \cos^2{\theta})  \cos{2\phi} \sin{2 \psi} + \cos{\theta} \sin{2 \phi} \cos{ 2 \psi}
\end{equation}
The amplitude of each lensed signal in the same signal sets will be different because the earth's rotation affects $\theta,\phi,\psi$ and each signal has a different magnification, the antenna pattern function of ET are a factor $\sin\gamma = \sqrt{3}/2$ smaller than above since the triangle shape.

If a network consists of $i$ detectors, the network's SNR will be defined as the square root of the sum of squares of each SNR of detectors:
\begin{equation}
\rho_\mathrm{network} = \sqrt{\rho^2_1+\rho^2_2+...+\rho^2_i},
\label{Eq:rho_network}
\end{equation}
It will be a good approximation even considered the confusion noise effect  \cite{Wu_2023}. In a detector network, we only need $\rho_\mathrm{network}$ upper to the identification threshold, which means a signal with SNR lower than the identification threshold can also be identified if the sum of squares of each SNR of detectors in a network upper to $\rho_0$.
In this paper, we consider the 3G detectors, which correspond Einstein Telescope alone (ET) and Cosmic Explorer, Einstein Telescope network (CEET). Table \ref{tab:lensed} shows the predictions of yearly lensed GW detection rates. 
As a comparison, we also show the results of the second generation detectors, i.e., LIGO Handford, LIGO Livingston, Advanced Virgo network (HLV). We can only identified signals with both $I_-$ and $I_{+}$ images being magnified above the threshold. Other situations such as only $I_-$ or $I_{+}$ being magnified above the threshold are not included in our statistics. The simulation result shows using 2G network (HLV) we can only detect about one SLGW event in a decade. Therefore, it is reasonable that no gravitational wave signal has been detected so far. The detection rate will rise by 2000 times using 3G detector (ET), we can detect 200.44 SLGW events per year. 3G detector network (CEET) will further improve detection rate to 300.38 per year, so 3G detectors are necessary for the study of SLGWs. These are consistent with previous results of \citet{10.1093/mnras/stab3298}. 
\begin{table}
\centering
\caption{Predictions of yearly lensed GW detection rates for both $I_-$ and $I_{+}$ images are magnified above the threshold under the 2nd, and 3rd generation detectors scenarios. Results are shown for the standard model of CBC formation in low-end metallicity scenario.
The mass distribution of BBH follows power-law with $\alpha=-2.3$ in the mass range 5-50 $M_{\sun}$}

\begin{tabular}{lccccccr}
\hline
\hline
CBC type & HLV & ET & ET+CE \\
\hline
BBH & 0.11/yr & 200.44/yr & 301.38/yr\\
\hline
\label{tab:lensed}
\end{tabular}
\end{table}

Then, preparing for identification,we estimate parameters of GW signals. In this study, we perform it using {\tt Bilby}\citep{2019ApJS..241...27A} with the dynesty sampler. We inject and estimate all signals with IMRPhenomPv3 as template from {\tt LALSuite} \citep{lalsuite}. We generate stationary Gaussian noise from the detectors’ PSDs. For the ET, we use the ET-D PSD \citet{Punturo_2010_1,Punturo_2010_2}, and for the CE, we use the projected PSD from \citet{2019BAAS...51g..35R}.

\section{Bayesian statistical framework}
We use Lens bayes factors to identify two lensed signals via their parameter comparison. This section introduces the bayesian statistical theory about Lens bayes factors concerning two aspects: the definition and numerical calculation of Lens Bayes factors, the relation between signal-to-noise ratios and Lens Bayes factors.

\subsection{Lens Bayes factor}
\label{sec:Lens Bayes factor}
\citet{2018arXiv180707062H} defined the Lens Bayes factor (LBF) as:
\begin{equation}
\Blu \equiv  \frac{\Zl}{\Zu} = \int d \vtheta~ \frac{\Pr(\vtheta|d_1)~\Pr(\vtheta|d_2)}{\Pr(\vtheta)}~,
\label{eq:lensing_bayes_factor}
\end{equation}
$\Blu$ show in which level that data supports lensed hypothesis. $\Zl$ is the marginal likelihood under the lensed hypothesis, $\Zu$ is the marginal likelihood under the unlensed hypothesis. $\Pr(\vtheta)$ is the prior.

Generally, parameter estimation will provide us with the posteriors $\Pr(\vtheta|d_1)$ and $\Pr(\vtheta|d_2)$. The parameters $\vtheta$ include the chirp mass $M_{c}$, mass ratio $M_{r}$, the dimensionless spin magnitudes $a_{1,2}$, the sky location RA, DEC, luminosity distance $D_\mathrm{L}$, coalescence phase $\phi$ geocent time $t_c$, and the orbital inclination $\theta_{JN}$ and $\psi$.
Figure \ref{fig:estimate_comparison.png}, for instance, shows distributions of marginalized posteriors of masses from typical BBH merger events.   

\begin{figure*}
	\centering
	\begin{minipage}{8 cm}
	
		\includegraphics[width=8 cm]{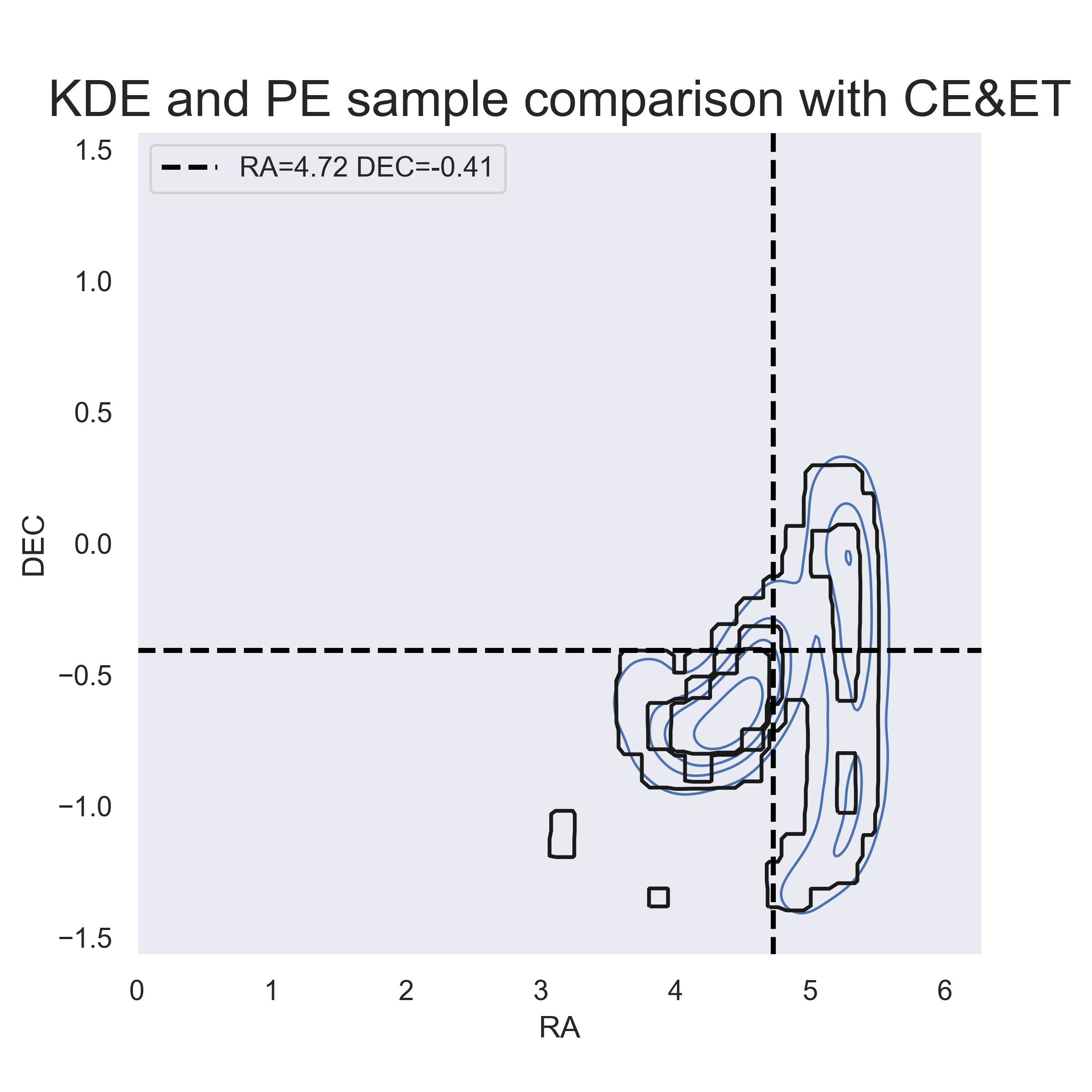}	
	\end{minipage}
	\begin{minipage}{8 cm}
	
		\includegraphics[width=8 cm]{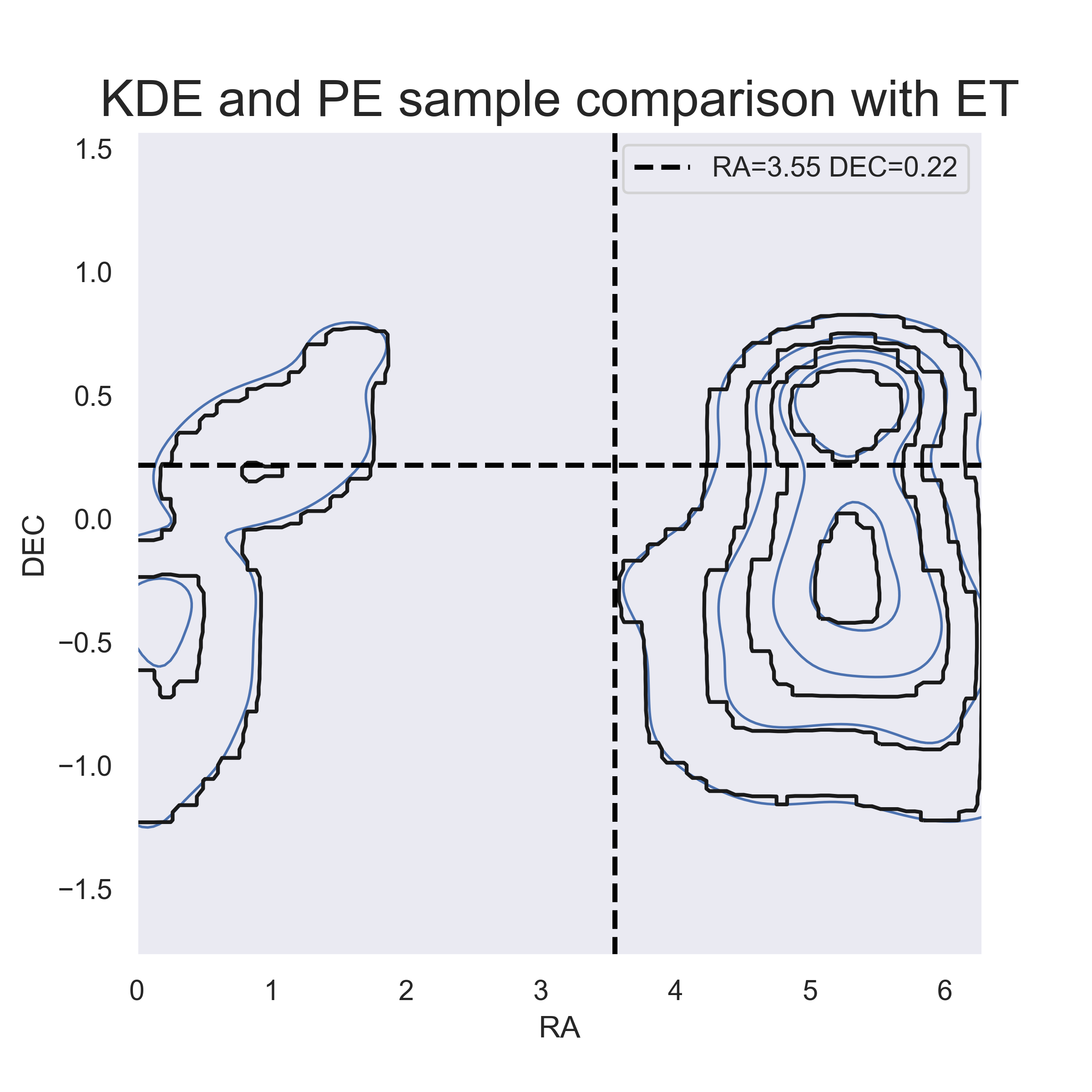}
	\end{minipage}
\caption{Examples of contour comparisons between PE results and results reproduced from KDE function. The SNR of the signal detected by CEET (left) is 10.7, the SNR of the signal detected by ET (right) is 9.1. The black contours are produced by KDE function. The blue contours are produced by PE. The dotted lines show the injection parameters.}
\label{fig:KDE—sample}
\end{figure*}

However, $D_\mathrm{L}$, $\phi$ and $t_c$ will be biased due to the unknown magnifications and different time delays, so we only consider $M_{c}, M_{r}, a_{1,2}$, RA, DEC, $\theta_{JN}$, $\psi$ and marginalize other parameters. For obtaining a continuous posterior distribution function from discrete sampling results, we perform kernel density estimations to the results of sampling. When we deal with low SNR signals, we find that their parameters are poorly limited, especially the spatial orientation. Due to the high dimension of the parameters, we can only take a small number of grids on each dimension. So it is challenging to transform such distributions into functions. In this paper, for signals with high SNR, we take 20 grids for each dimension. For a part of signals with low SNR, we take 40 grids for each dimension. Figure \ref{fig:KDE—sample} shows a contour comparison between parameter estimation (PE) results and results reproduced from KDE function. We can see that KDE method can basically reproduce this kind of complex non Gaussian distribution. \cite{2021MNRAS.506.5430J} provide a fast methodology to analyse strongly lensed event pairs, which replaced the prior used in the analysis of one SLGW image by the posterior of another image and avoiding using of KDE. It will be a good crosscheck to our results. We will consider this method for analysing signals producing more than two images in the future.

We multiply the KDE of $\Pr(\vtheta|d_1)$, $\Pr(\vtheta|d_2)$ and integrat them by using numerical integration through Monte Carlo method \citep{Lepage_2021}.  To improve the accuracy of KDE further, We divide the $\vtheta$ into intrinsic parameters $\vtheta_{intr}(M_c,M_r,a_1,a_2)$ and external parameters $\vtheta_{ex}(ra,dec,\psi,\vtheta_{JN})$. We perform the KDE separately and multiply them as the whole posterior density function.

\subsection{The analytical approximate relationship between SNR and LBF}
\label{sec:The analytical approximate relation between SNR and LBF}
The Lens Bayes factor (LBF) is different from the Bayes factor of one signal. 
Since we need at least two signals to perform integral, we cannot directly follow the treatment in \cite{Cornish_2011} or \citep{2021PhRvD.103j4055W}, which considered the relationship between SNR and Bayes factor under different assumptions of one signal to find the analytical relationship between SNRs and Bayes factors. 
However, we can obtain the relationship by expanding parameter space.

We treat the LBF as:  
\begin{equation}
\begin{split}
\Blu :&= \frac{\Zl}{\Zu}=\frac{\Pr(d_1,d_2|L)}{\Pr(d_1|U) \Pr(d_2|U)}~\\&=\frac{\int d \vtheta  \mathcal{L}_1 (\vtheta|L)\mathcal{L}_2 (\vtheta|L)\pi(\vtheta|L)}{\int d \vtheta\mathcal{L}_1 (\vtheta|U)\pi(\vtheta|U)\cdot \int d \vtheta\mathcal{L}_2 (\vtheta|U)\pi(\vtheta|U)},
\label{eq:1}
\end{split}
\end{equation}
where $L$ is the lensed hypothesis, $U$ means the unlensed hypothesis,
$\mathcal{L}_{1,2}(\vtheta)\equiv \Pr(d_{1,2}|\vtheta)$ is the likelihood function. $\pi(\vtheta)$ is the prior function. 

For denominator of  Eq. (\ref{eq:1}), we want to combine the two integral expressions into one. So we define $\Theta\equiv \vtheta_1 \otimes \vtheta_2$, where $\vtheta_1$ and $\vtheta_2$ are from the identical parameter space of single signal. Subscripts 1,2 are used to distinguish them. Then, we take $|\theta\rangle = \frac{1}{c}|h\rangle $ as normalized GW signal corresponding to parameter $\theta$, and $|\mathbf{h}\rangle \equiv |h_1\rangle \otimes |h_2\rangle$, $|\mathbf{d}\rangle \equiv |d_1\rangle \otimes |d_2\rangle$. Thus, the denominator can be written as 
\begin{equation}
\int d {\Theta}  \mathcal{L}_1 (\vtheta_1|U)\mathcal{L}_2 (\vtheta_2|U)\pi({\Theta}|U).
\label{2_1}
\end{equation}
Similarly the numerator, 
\begin{equation}
  \int d {\Theta}  \mathcal{L}_1 (\vtheta_1|L) \mathcal{L}_2 (\vtheta_2|L)\delta (\vtheta_1-\vtheta_2)\pi({\Theta}|L).
\label{2_2}
\end{equation}
where $\delta (\vtheta_1-\vtheta_2)$ is Dirac delta function. Notice that prior functions are the same in both lensed and unlensed hypothesis, and $\vtheta_1=\vtheta_2$ under lensed hypothesis. Combining $\mathcal{L}_1$and$\mathcal{L}_2$, $\tilde{{\mathcal{L}}}(\Theta|U)= \mathcal{L}_1 (\vtheta_1|U)\mathcal{L}_2 (\vtheta_2|U)$, $\tilde{{\mathcal{L}}}(\Theta|L)= \mathcal{L}_1 ({\vtheta_1}|L)\mathcal{L}_2 ({\vtheta}_2|L)\delta (\vtheta_1-\vtheta_2)$.

Then, Eq. (\ref{eq:1}) takes form as:
\begin{equation}
\label{Eq:3}
\Blu =  \frac{\int d\Theta \tilde{{\mathcal{L}}}(\Theta|L) \pi(\Theta|L)}{\int d\Theta \tilde{\mathcal{L}}(\Theta|U) \pi(\Theta|U)}.
\end{equation}

We notice that $\tilde{{\mathcal{L}}}(\Theta|U)$ and $\tilde{{\mathcal{L}}}(\Theta|L)$ are still gaussian functions. We can treat $\Blu$ similar as the type II image bayes factor developed \citet{2021PhRvD.103j4055W}. According to the result of them, we can write the log LBF as :
\begin{equation}
\label{Eq:4}
\begin{aligned}
	\ln \Blu \approx&	\ln \frac{\tilde{{\mathcal{L}}}(\Theta^{\mathrm{MLE}}|L)}{\tilde{{\mathcal{L}}}(\Theta^{\mathrm{MLE}}|U)}+\ln  \frac{\Sigma^{\rm posterior}_\mathrm{L}}{\Sigma^{\rm posterior}_{U}},
	\end{aligned}
\end{equation}
where $\Theta^{\mathrm{MLE}}$ is the Maximum Likelihood Estimation of $\Theta$, $\Theta^{\mathrm{MLE}}=\vtheta^{\mathrm{MLE}}_1\otimes\vtheta^{\mathrm{MLE}}_2$. $\Sigma^{\rm posterior}$ is joint posterior volume. Thus, according to the previous definition, we can split Eq. (\ref{Eq:4}) as:
\begin{equation}
\label{Eq:5}
\begin{split}
\ln \frac{\tilde{{\mathcal{L}}}(\Theta^{\mathrm{MLE}}|L)}{\tilde{{\mathcal{L}}}(\Theta^{\mathrm{MLE}}|U)} = \ln\frac{\mathcal{L}_1(\vtheta^{\mathrm{MLE}}_1|L)}{\mathcal{L}_1(\vtheta^{\mathrm{MLE}}_1|U)} +	\ln\frac{\mathcal{L}_2(\vtheta^{\mathrm{MLE}}_2|L)}{\mathcal{L}_2(\vtheta^{\mathrm{MLE}}_2|U)}\\+\ln \delta (\vtheta^{\mathrm{MLE}}_1-\vtheta^{\mathrm{MLE}}_2).
\end{split}
\end{equation}
Theoretically, for lensed GW signals, we have $\Theta_{\mathrm{MLE}}=\vtheta_{\mathrm{MLE}}\otimes\vtheta_{\mathrm{MLE}} \to \vtheta_{\mathrm{true}}\otimes\vtheta_{\mathrm{true}}$, 
$(\vtheta^{\mathrm{MLE}}_1-\vtheta^{\mathrm{MLE}}_2)\to 0 $.  
According to \citep{2021PhRvD.103j4055W}, we have 
\begin{equation}
\label{Eq:6}
\begin{aligned}
\ln \frac{{{\mathcal{L}_i}}(\vtheta|L)}{{{\mathcal{L}_i}}(\vtheta|U)} \approx \epsilon_i \rho_i^2
	\end{aligned}
\end{equation}
For second term of Eq. (\ref{Eq:4}), under unlensed hypothesis, we can estimate the posterior volume $\sigma^{\rm posterior}_{i}$ in direct product space as 
\begin{equation}
\label{Eq:7}
\begin{split}    
\Sigma^{\rm posterior}_{U}=\sigma^{\rm posterior}_{1} \times \sigma^{\rm posterior}_{2} 
\end{split}
\end{equation}
where we can roughly get:
\begin{equation}
\label{Eq:posterior}
		\sigma^{\rm posterior}_{i} \approx \prod_{j=1}^{N} \sqrt{2\pi} \Delta \theta^{j, \;{\rm posterior}}_{i}.
\end{equation}
Under lensed hypothesis, the situation is complicated. Since we have ensured $\vtheta_1= \vtheta_2$ in simulated SLGW pairs, the posterior volume in direct product space should be
\begin{equation}
\label{Eq:lens_posterior}
\begin{aligned}
\ln\Sigma^{\rm posterior}_{L}&\approx\ln\sigma^{\rm posterior}_{L}/\delta(\vtheta^{\mathrm{MLE}}_1-\vtheta^{\mathrm{MLE}}_2)\\
&=\ln \sigma^{\rm posterior}_{L} -\ln \delta(\vtheta^{MLE}_1-\vtheta^{\mathrm{\mathrm{MLE}}}_2)
\end{aligned}
\end{equation}
Substituting \ref{Eq:4} with \ref{Eq:5},\ref{Eq:6}, \ref{Eq:7} and \ref{Eq:lens_posterior}, we get rid of infinite term $\delta(\vtheta_1^{MLE}-\vtheta_2^{\mathrm{MLE}})$ 
\begin{equation}
\label{Eq:fanal_1}
\begin{aligned}
\ln \Blu \approx \epsilon_1\rho_1^2 + \epsilon_2\rho_2^2 +\ln\frac{\sigma_\mathrm{L}^{\rm posterior}}{\sigma_1^{\rm posterior}\sigma_2^{\rm posterior}} 
	\end{aligned}
\end{equation}
where $\epsilon$ is the mismatch factor. $1-\mathbf{\epsilon_i}=\langle\theta_\mathrm{L}|\theta_i\rangle$. Note that the gravitational lensing effect will not change the waveform of GWs, and $|d_{1,2}\rangle$ are simulated under the same parameters and we only consider unbiased parameters showed in \ref{sec:Lens Bayes factor}, we have $\langle\theta_\mathrm{L}|\theta_i\rangle=1$ and $\epsilon_{1,2}=0$. Second, $\Delta \theta^{j, \;{\rm posterior}}_{i} \propto 1/\sqrt{\langle h_{i} | h_{i} \rangle} \propto 1/\rho_i$. For calculating $\sigma_\mathrm{L}^{\rm posterior}$, since we only consider the unbiased parameters, two lensed GWs arriving earth with a time gap equal to one signal detected by two different detector networks jointly. Considering Eq. (\ref{Eq:rho_network}) and (\ref{Eq:posterior}), we have $\sigma_\mathrm{L}^{\rm posterior}\propto {\frac{1} {\sqrt{\rho_1^2+\rho_2^2}}}^N$ and $\sigma_i^{\rm posterior}\propto [{\frac{1} {\rho_i}}]^N$.

Eventually, we get the analytic relationship between LBF and $\rho_{1,2}$:
\begin{equation}
\label{Eq:snr_bf}
\begin{aligned}
\ln \Blu \approx  {N}\ln(\rho)+b,
	\end{aligned}
\end{equation}
where 
\begin{equation}
\label{Eq:rho}
\rho=\frac{\rho_1\rho_2} {\sqrt {\rho_1^2+\rho_2^2}}, 
\end{equation}
$b$ is consist of the log scale factor of Eq. (\ref{Eq:posterior}) and $\Delta \theta^{j, \;{\rm posterior}}_{i} \propto 1/\sqrt{\langle h_{i} | h_{i} \rangle} \propto 1/\rho_i$. It is a constant that depends on the waveform template. Note that $\rho_1$=$\rho_{I_-}$ and $\rho_2$=$\rho_{I_+}$ in SIS lensing model. It is different from the common quadratic function as \cite{Cornish_2011} leading by mismatch factor. 

Generally, we can find that LBFs are positively correlated with SNRs. 
Eq. \ref{Eq:snr_bf} indicate for a pair of lensed GWs, the more consistent parameters we take into account, the higher $\Blu$ we can get. This is the analytic interpretation of FIG. 6 of \cite{2018arXiv180707062H}. We will give more details in Section \ref{sec:SNR and LBF}.  
Signal pairs with extreme magnification ratios ($\mu_-/ \mu_+ < 0.1$) are an example of a special case. In these cases $\rho\approx\rho_-$, which means $\Blu$ or the identification capability mainly depends on the fainter signal in these cases. 

The relationship Eq. \ref{Eq:snr_bf} are still valid to the multiple imaging events that we get more than two lensed GW signals. We will discuss it in the future. 
\section{RESULTS AND DISCUSSION}
\subsection{The LBF under the lensed and unlensed hypotheses}

\begin{figure*}
    \centering
    \includegraphics[width=2\columnwidth]{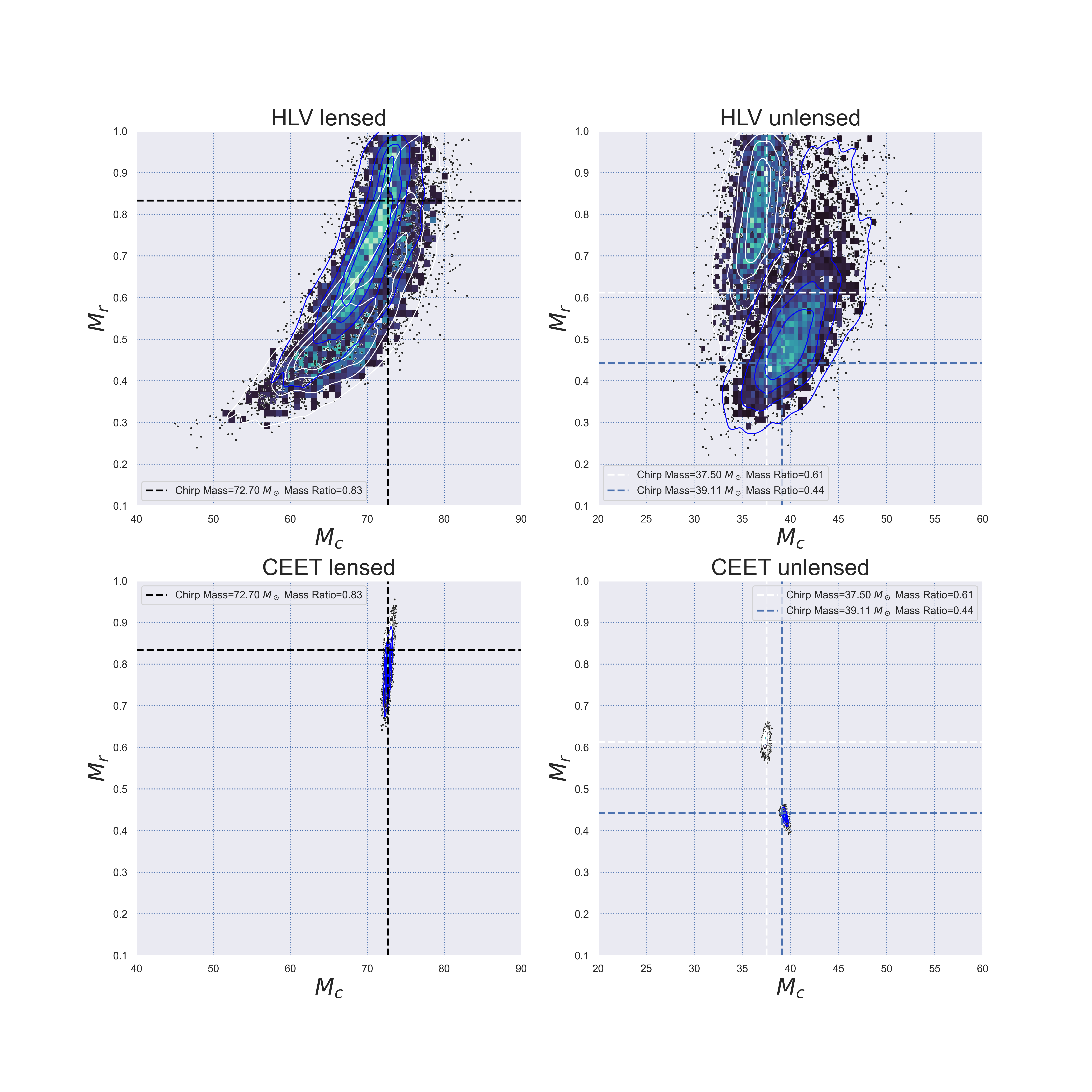}
    \caption{Marginalized posteriors of chirp mass ($M_c$) and mass ratio ($M_r$) of simulated GW signals. The posterior distribution in the upper left image is two signals based on HLV network under the lensed hypothesis. The upper right is based on HLV network under the unlensed hypothesis. The two images below show the posterior distributions of GW signals with the same injection parameters and hypotheses as the two images upper, but based on the CEET network. The dotted lines show the injection parameters.}
    \label{fig:estimate_comparison.png}
\end{figure*}

In Figure \ref{fig:estimate_comparison.png}, we compare the estimation results of pairs of typical BBH signals from LIGO Handford, LIGO Livingston, Advanced Virgo network (HLV) and 3G detector network (CEET) under the lensed or unlensed hypothesis. It is clear that using CEET will substantially increase the accuracy of parameter estimation. For the lensed signal pairs, the better constrained distributions makes LBF larger as shown in the left two figures of Fig. \ref{fig:estimate_comparison.png}, which in fact is the intuitive expression of the Eq. (\ref{Eq:snr_bf}). For the unlensed pairs, the `false' overlap that causes by two independent event pairs happening to be similar will be more unlikely as shown in the two right figures of Fig. \ref{fig:estimate_comparison.png}. As a result, using the 3G detector network to identify SLGWs will be more effective. 

\begin{figure}
	\centering
	\begin{minipage}{\linewidth}
		\centering
		\includegraphics[width=\linewidth]{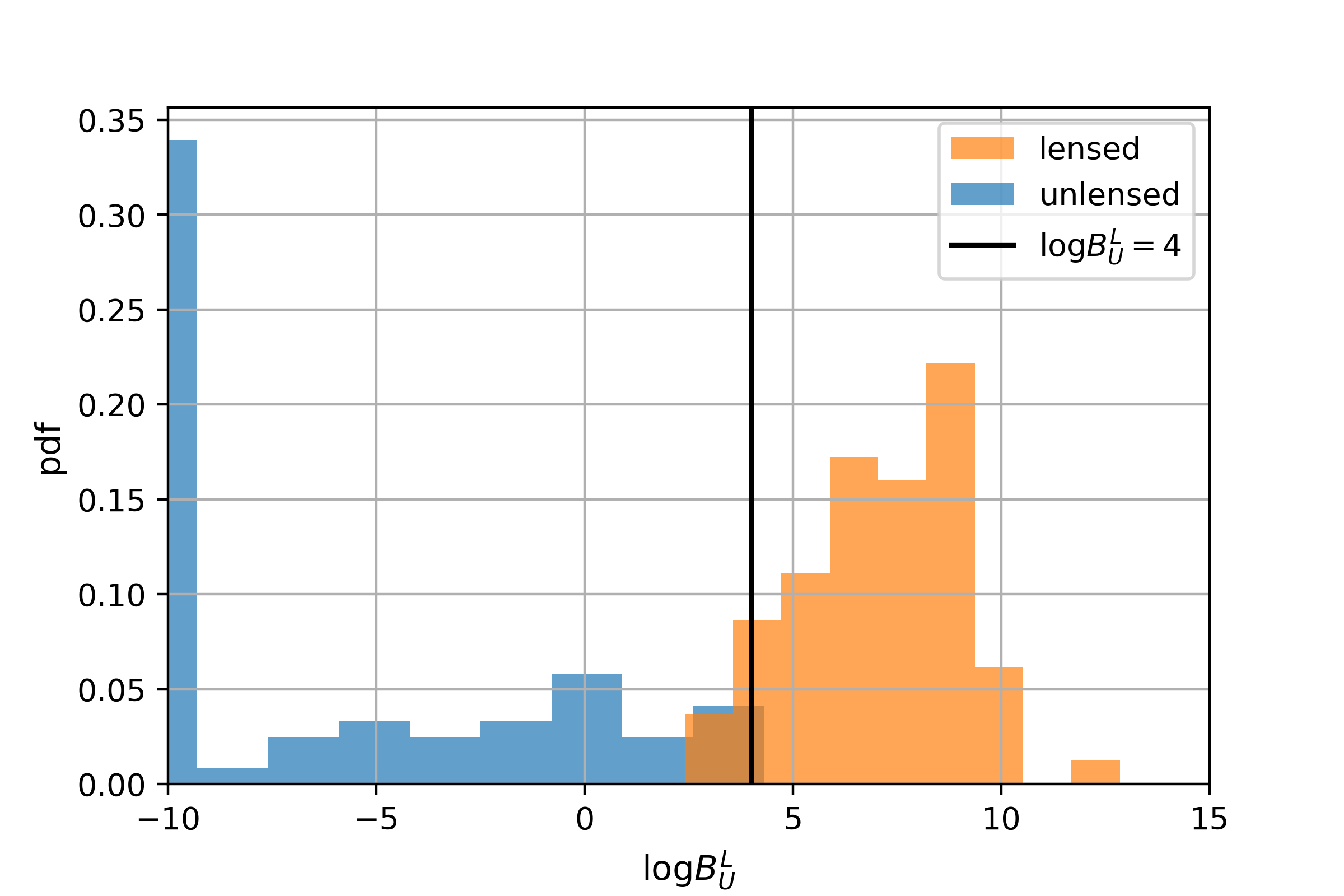}	
	\end{minipage}
	
	\begin{minipage}{\linewidth}
		\centering
		\includegraphics[width=\linewidth]{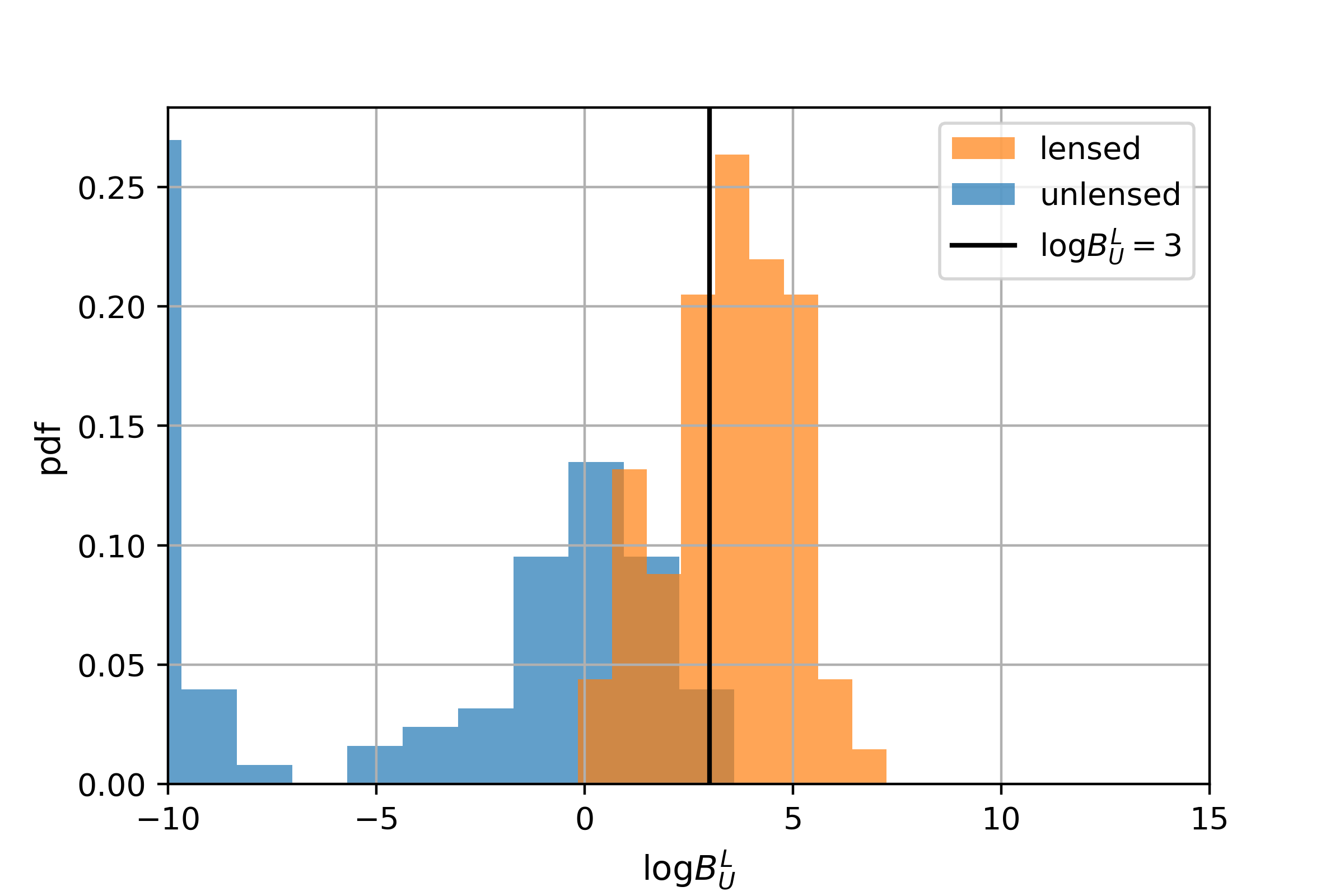}
	\end{minipage}
\caption{Distribution of the $log_{10}\Blu$ computed from the unlensed and lensed pairs detected jointly by CE\&ET (upper) and ET alone (lower). The selected threshold value is represented by a black line.}
\label{bays_factor_both.png}
\end{figure}

Figure \ref{bays_factor_both.png} shows the distributions of $log_{10}\Blu$ of lensed and unlensed event pairs under 3rd generation network detection scenario. The Lens Bayes factors are computed by multiplying the 8-dimensional kernel density estimates of posterior distributions of $I_-$ and $I_+$, which include $(M_c,M_r,a_1,a_2,ra,dec,\psi,\theta_{JN})$, and integrating them (\ref{eq:lensing_bayes_factor}). We can see that the distributions of LBFs under lensed and unlensed hypotheses are so different that we can choose some values as discriminators to identify SLGW pairs. However, it is important to note that unlensed signal pairs in Figure \ref{bays_factor_both.png} were not selected at random because almost all unlensed pairs will give us very small LBFs ($\Blu<10^{-10}$). In other words, the higher accuracy of parameter estimation, the smaller probability that two independent pairs' parameter space overlap. So we need not calculate LBFs for pairs whose differences of injection parameters are relatively large enough. Instead, for unlensed events of CEET, considering the average of posterior (uncertainty) volume, we limit the difference of true value between each event of pairs as Table \ref{tab:limit} and seek out the potential pairs. For other event pairs, we assume their $\Blu$ are smaller than -10 directly. For ET alone, however, we cannot limit the spatial location (RA, DEC) because it is difficult to locate BBH signals with short duration by only one detector. So we keep RA and DEC unlimited and limit others. According to the integral results Fig. \ref{bays_factor_both.png}, we take $log_{10} \Blu >4$ for CEET as the threshold with false alarm rate as about $10^{-6}$; we take $log_{10} \Blu >3$ for ET as the threshold with false alarm rate as about $10^{-5}$. These false alarm rates are consistent with the result of \cite{_al_kan_2023}, which provided many details about the false alarm rate of lensed events.
\begin{table}

\centering
\caption{The limit of injection parameters we choose to find unlensed signal pairs.}

\begin{tabular}{lccccc}
\hline
\hline
parameters & unit & limit \\
\hline
$m_1^z$   & $M_{\sun}$ & $\Delta m_1^{\mathrm{s}}<0.1m_1^{\mathrm{s}}$ \\
$m_2^{\mathrm{s}}$   & $M_{\sun}$ & $\Delta m_2^{\mathrm{s}}<0.1m_2^{\mathrm{s}}$ \\

$a_{1}$ (BBH) & 1 & $\Delta a_{1} <1$ \\
$a_{2}$ (BBH) & 1 & $\Delta a_{2} <1$ \\
RA (only for CEET) &  rad. &  $\Delta$ RA <0.5      \\
DEC (only for CEET) &  rad. &  $\Delta$ DEC<0.5      \\
$\theta_{JN}$ & rad. & $\Delta\theta_{JN} <0.5$ \\
$\psi$ & rad. & ---- \\
\hline
\label{tab:limit}
\end{tabular}
\end{table}

\subsection{SNR and LBF}
\label{sec:SNR and LBF}

\begin{figure}
	\centering
	\begin{minipage}{\linewidth}
		\centering
		\includegraphics[width=\linewidth]{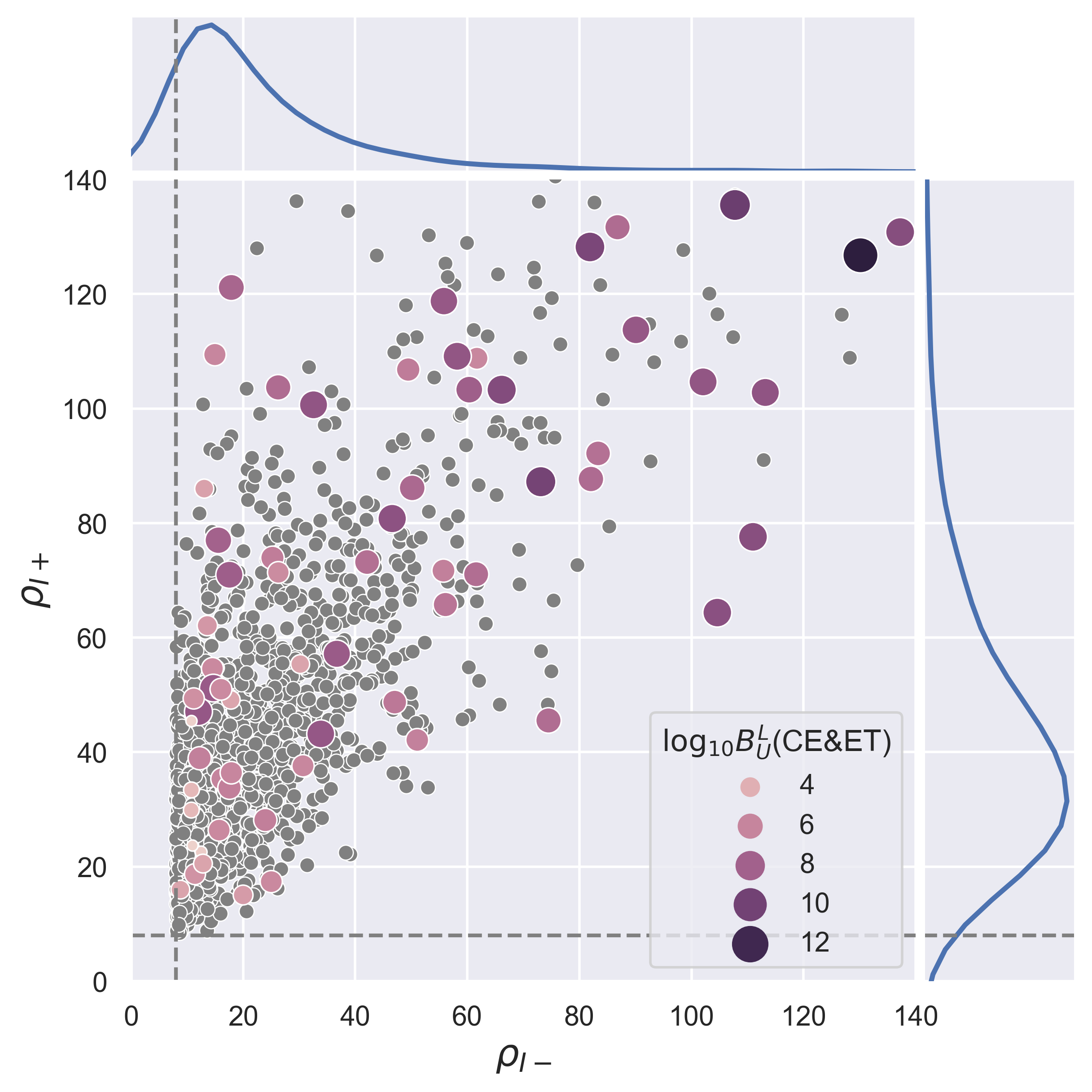}	
	\end{minipage}

	\begin{minipage}{\linewidth}
		\centering
		\includegraphics[width=\linewidth]{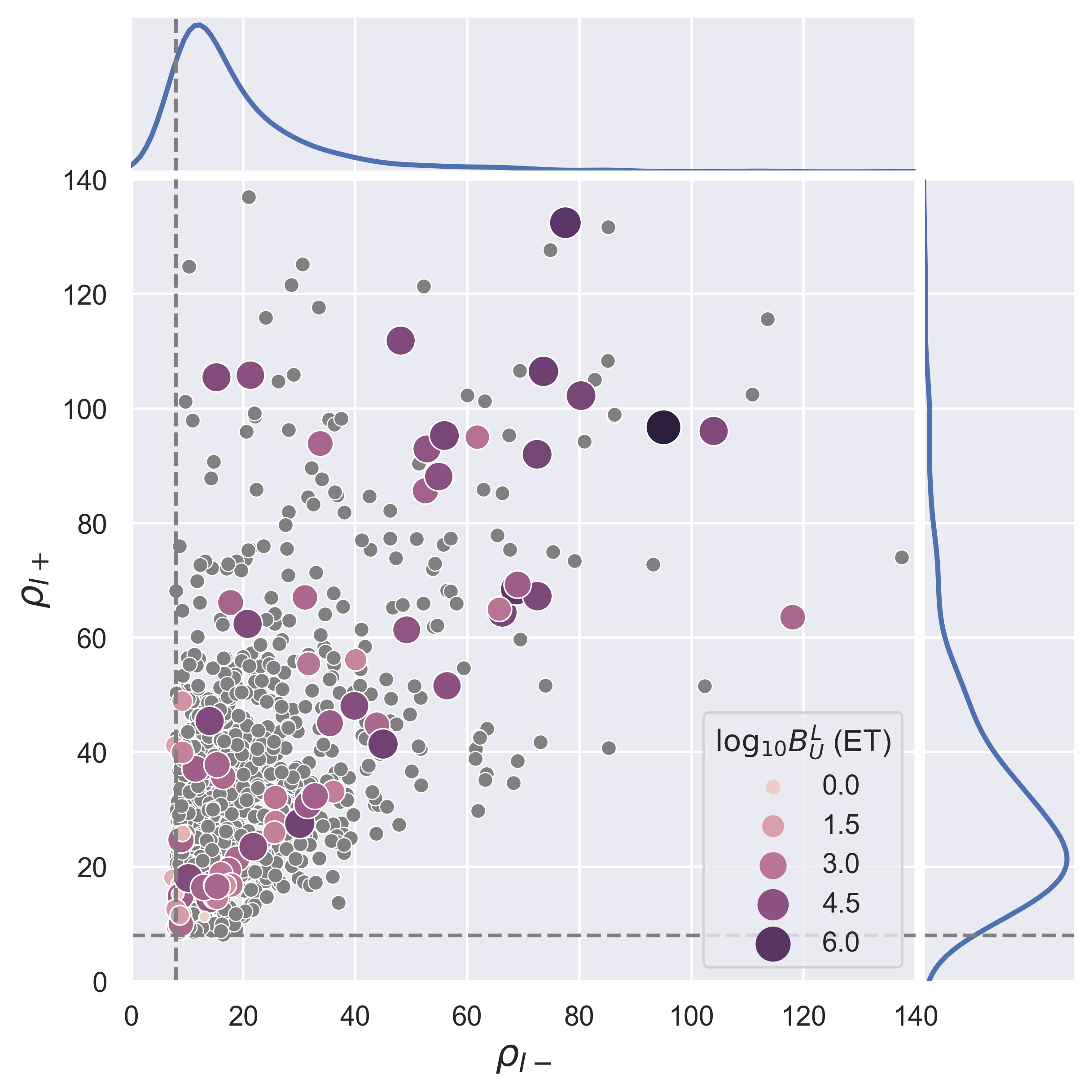}
	\end{minipage}
    \caption{SNR distribution of SLGW signals and Bayes factors of CEET. Each grey point represent the lensed pairs from BBH. Points with bigger sizes and colors from light to dark range represent the selected lensed pairs. The darker their colors, the greater their Bayes factors. The dashed line shows the detector’s detection threshold, we set it at 8.}
    \label{SNR_comparison.png}
\end{figure}

\begin{figure}
    \centering
    \includegraphics[trim=1mm 1mm 1mm 1mm, clip, width=80mm]{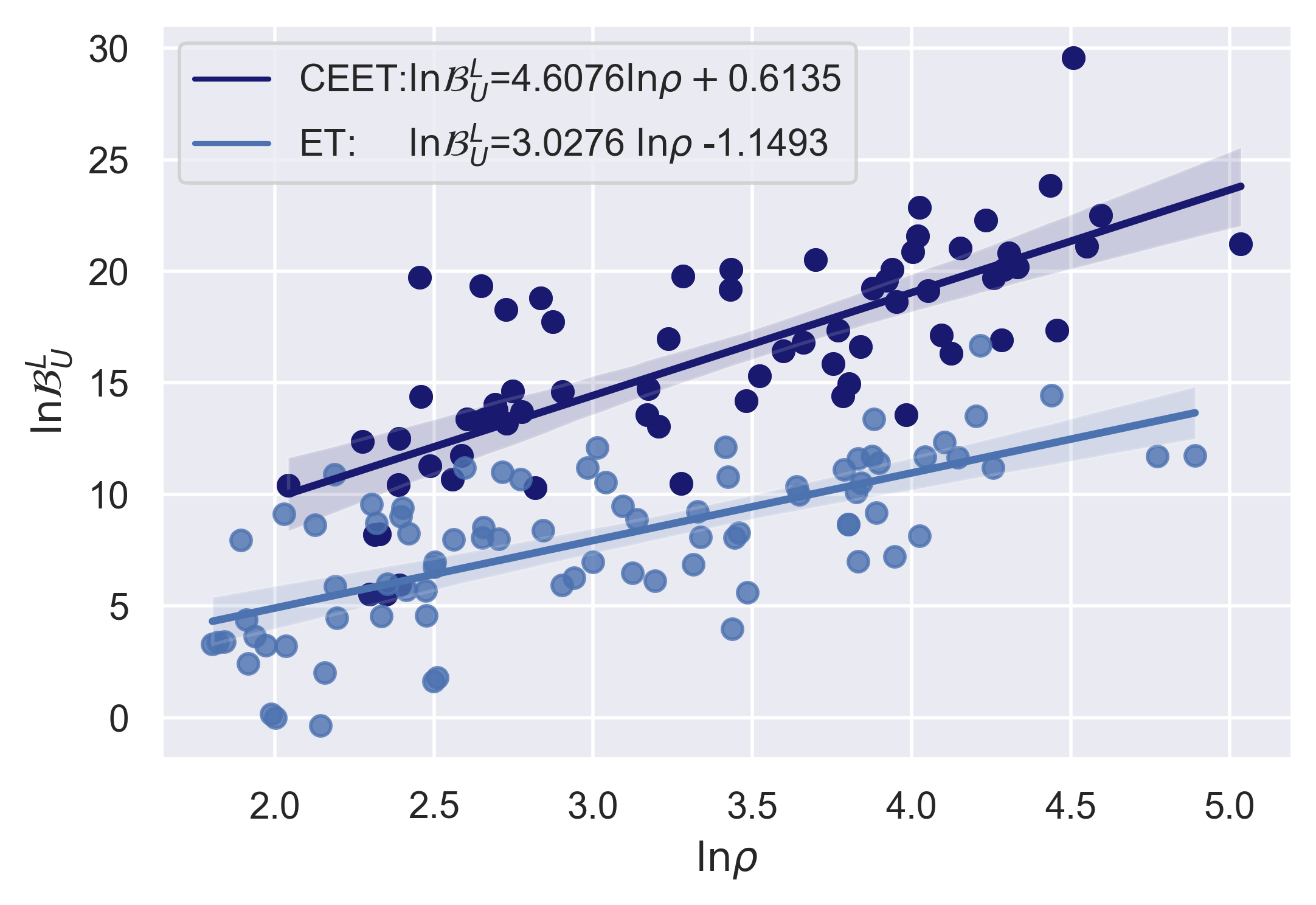}
    \caption{The relationship between SNR and LBF of simulated lensed events detected by ET alone and CEET jointly, $\rho\equiv\frac{\rho_1\rho_2} {\sqrt {\rho_1^2+\rho_2^2}}$.}
    \label{bf_and_SNR.png}
\end{figure}

Figure \ref{SNR_comparison.png} shows the SNR distribution of SLGW signals and their LBFs. We girded the SNR distribution, each grid size is $20\times 20$. Then we randomly select lensed pairs in each grid to get $\Blu$ as representative values of that range. It is clear that LBFs are positively correlated with SNRs. Based on our predictions, 200-400 lensed GW events would be registered by CEET network. Unfortunately, 42.5\% of signals are concentrated in $8< \rho_{I_+} < 40$, $8< \rho_{I_-} < 20$. Within this range, $log_{10}\Blu$ values range from 3 to 6. Although most of log LBFs are higher than 4, there are still 30\% of signal pairs lower than 4, which cannot be identified. For pairs with higher SNRs, almost all of them can be identified clearly. Considering the distribution of signals, 87.3\% of the previous prediction results \cite{10.1093/mnras/stab3298} can be identified, that is 263/yr with with a low false alarm rate of $10^{-6}$. It is still far more efficient than using advanced LIGO and Virgo, which could identifies $10\sim15\%$ of the lensed event pairs with a false alarm probability of $10^{-5}$ if we only calculate the LBF \citep{2018arXiv180707062H}. 

For ET alone, however, the result is disappointing. Because of lacking of other detectors with similar sensitivity, signals detected by ET alone have low spatial resolution. This is a obstacle. Like earlier studies, the source sky location parameters (RA, DEC) are the ones that primarily enhance the LBF's performance. \citep{2018arXiv180707062H}, the poor spatial location ability causes only 50.6\% of lensed signal pairs can be identified, and the relationship between SNR and LBF is not clear neither. Considering the previous prediction results, only 95.14/yr detected can be identified. As a result, it is clear that joint detection using detector networks is essential to identifying lensed events. This is true not only because it lowers the signal threshold at a single detector to enable the detection of more GW events, but also because it improves the spatial resolution to provide better constrained posterior distributions, which increase the performance of the LBF.

\begin{figure}
    \centering
    \includegraphics[trim=1mm 1mm 1mm 1mm, clip, width=80mm]{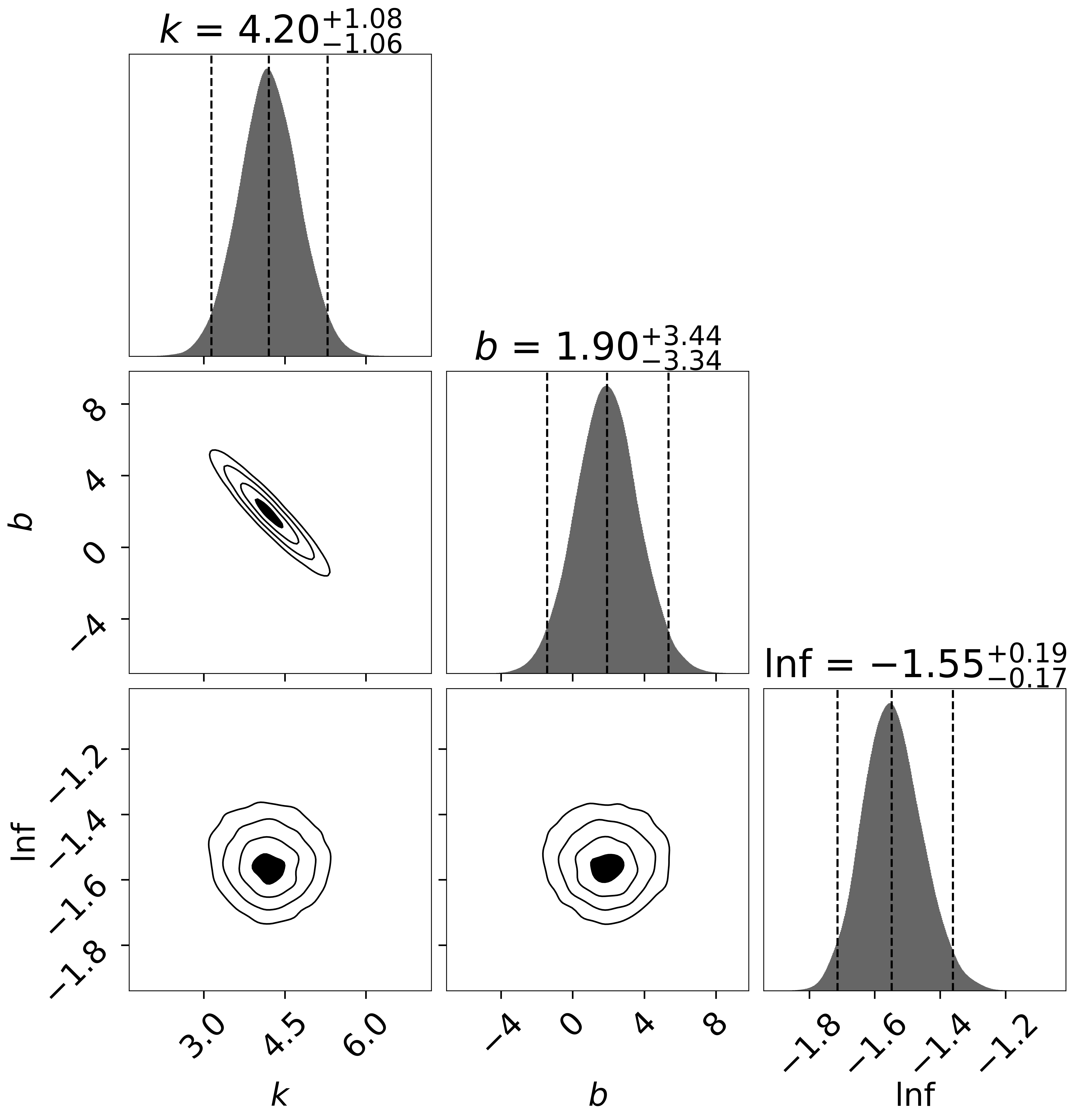}
    \caption{Linear regression diagram fitting with $ \ln \Blu=k{\ln(\frac{\rho_1\rho_2} {\sqrt {\rho_1^2+\rho_2^2}})}+b$ by Bayes parameter estimation for CEET with log-likelihood for Gaussian distribution.}
    \label{mcmc.png}
\end{figure}
\begin{figure}
    \centering
    \includegraphics[trim=1mm 1mm 1mm 1mm, clip, width=80mm]{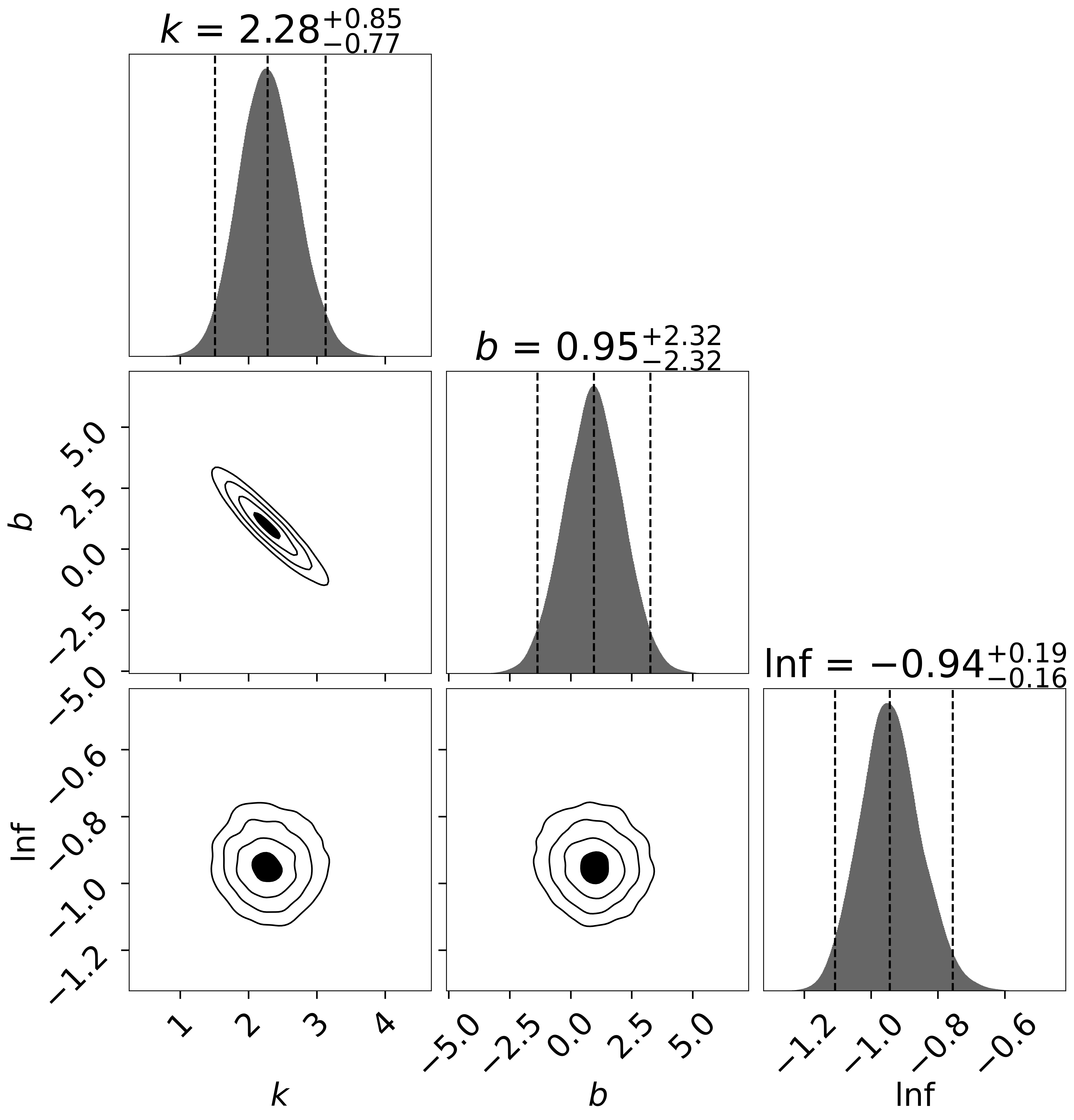}
    \caption{Linear Regression diagram fitting with $ \ln \Blu=k{\ln(\frac{\rho_1\rho_2} {\sqrt {\rho_1^2+\rho_2^2}})}+b$ by Bayes parameter estimation for ET with log-likelihood for Gaussian distribution.}
    \label{mcmc_ET.png}
\end{figure}
Figure \ref{bf_and_SNR.png} show the Bayes factors of lensed events with different SNRs. LBFs are positively correlated with SNRs. 
According to Eq. (\ref{bf_and_SNR.png}), gradient k is equal to $4.20^{+1.08}_{-1.08}$ for CEET joint detection, and $2.28^{+0.85}_{-0.77}$ for ET alone. Theoretically, if we ignore the correlation between each parameter and all posterior distributions of considered parameters are unbiased Gaussian distributions, we have $k=N=8$ because we use 8 parameters in total. But these presumptions are not match the actual situation.
The first and most important reason is some parameters such as $\psi,\theta_{JN}$ are estimated poorly in general. They are relatively invalid parameters in fact. Second, $M_c,M_r,a_1,a_2$ are highly relevant to each other, so do the RA and DEC. Therefor Eq. (\ref{Eq:posterior}) is a rough approximation. Third, the true background noise cannot be standard Gauss noise. All these will make $k$ bias away from $N$. Thus, we can define $N_{\mathrm{eff}}$ as `Number of parameters that can be well estimated' to indicate the effective parameters we choose, and Eq. (\ref{Eq:snr_bf}) should be $\ln \Blu = {N_{\mathrm{eff}}}\ln(\rho)+b$. For CEET, if we roughly take $\psi,\theta_{JN}$ as invalid parameters, we have $N_{\mathrm{eff}}\le 6$, the simulation result is $4.20^{+1.08}_{-1.06}$. For ET alone, if we take $\psi,\theta_{JN}$, RA, DEC as invalid parameters, we have $N_{\mathrm{eff}}\le 4$, the simulation result is $2.28^{+0.85}_{-0.77}$.

According to Fig. \ref{bf_and_SNR.png}, if we want to identify SLGW pairs by LBF, we should set threshold to the SNR as $\rho=\frac{\rho_1\rho_2} {\sqrt {\rho_1^2+\rho_2^2}}>6.5$ for CEET, and $\rho=\frac{\rho_1\rho_2} {\sqrt {\rho_1^2+\rho_2^2}}>14.3$ for ET alone.


\section{CONCLUSIONS AND OUTLOOK}
In this work, we calculated the Lens Bayes factor of simulated SLGW signal pairs and unlensed pairs with detector networks of the 3rd generation. We used the {\tt StarTrack} population synthesis code to estimate the intrinsic merger rate of BBHs. We used singular isothermal sphere (SIS) model to simplify the lensed GW signal statistics.
We used {\tt Bilby} with the {\tt dynesty} sampler to estimate parameters of GW signals. Then we calculated the Lens Bayes factor (LBF) of GW pairs by integrating their posterior density function.
We determined the LBF thresholds to distinguish lensed signal pairs from unlensed signal pairs by comparing the LBF distribution of the two types of signal pairs. Through analytical derivation and simulation, we found out the relationship between SNR and LBF. Here are our conclusions.
 
1.The Lens Bayes factor is a good determiner to identify strongly lensed gravitational waves detected by third-generation detector networks. The 3G detectors not only can increase the detection rate of the lensed GWs but also can improve our ability of identification. For SLGW pairs detected by CEET, with SNR >8, there is a positive correlation between the identification ability and the SNR. For pairs with SNRs belong to $8< \rho_{I_+} < 40$, $8< \rho_{I_-} < 20$, 30\% of them cannot be identified. For pairs with higher SNRs, almost all of them can be identified clearly. Considering the distribution of signals, 87.3\% of the previous prediction results \cite{10.1093/mnras/stab3298} can be identified, which means about 263/year.
For SLGW pairs detected by ET with SNR >8, 50.6\% of them can be identified by Lens Bayes factors. Considering the previous prediction results, only 100/year signal pairs can be identified. Poor restrictions on RA and DEC are the main reason for this result. Consequently, it is crucial to identify lensed events using joint detection with detector networks.

2. For lensed pairs, higher SNRs will be benefit to identify them. For unlensed pairs, higher SNRs of unlensed pairs will reduce the false alarm rate by shrinking the posterior space. According to Figure \ref{bf_and_SNR.png}, if we set $ \log_{10}\Blu=4$ for CE\&ET and $\log_{10}\Blu=3$ for ET as determining thresholds, we should set SNR threshold as $\rho=\frac{\rho_1\rho_2} {\sqrt {\rho_1^2+\rho_2^2}}>6.5$ for CEET, and $\rho=\frac{\rho_1\rho_2} {\sqrt {\rho_1^2+\rho_2^2}}>14.3$ for ET alone. It means for a pair of `potential lensed signals', LBF is averagely effective only if their signal-to-noise ratios are large enough. Particularly, for a signal pair detected by CE\&ET, if their SNRs are close, we need $\rho_1, \rho_2 \gid 9.2$. If there is a large gap between their SNRs, in other words, if the angular position of the source is close to Einstein radius, the fainter one should be larger than $6.5$. For a signal pair detected by ET alone, if their SNRs are close, we need $\rho_1, \rho_2 \gid 20.2$. If there is a large gap between their SNRs, the lower one should be larger than $14.3$.    

In fact, we just simply calculated the least overlapping parameters of the lensed GWs. If we consider other lens properties especially the image type, such as the Hilbert transform between type I and type II images, the phase shift between type I and type III images, the identification efficiency will be improved. There are also other techniques to identify SLGWs, such as using coherence ratio with the joint-estimate method. Including selection effects and astrophysical information, they can find out the more realistic Bayes factor $\Blu$. These can be done by the {\tt LALInference}-based pipeline \cite{Liu_2021} and {\tt hanabi} pipeline \cite{2021arXiv210409339L}. 
For a single signal (corresponding to the specific lensed image), the waveform would be slightly distorted by lensing, so we can play on this property to calculate Bayes factor as  \cite{2021PhRvD.103j4055W}, \cite{2020arXiv200712709D}, and \cite{Liu_2021}.  
We also notice that \cite{2021PhRvD.104j3529D} penalize the use of SIS model in the \citet{Abbott_2021}, they think the ellipticity plays an important role in researches about lensed GWs. We can consider this in future work. 
It is worth applying these techniques to 3G detector network in the future. But we need to point out all these techniques belong to the consistency test and depend on the overlap of parameters, so our work is simple but valuable.

\section*{Acknowledgements}
KL was supported by National Natural Science Foundation of China (NSFC) No. 12222302, 11973034 and Funds for the Central Universities (Wuhan University 1302/600460081).
LY acknowledges support by JSPS KAKENHI Grant Number JP 21F21325.
ZHZ was supported
by NSFC Nos. 12021003, 11920101003, and 11633001 and the
Strategic Priority Research Program of the Chinese Academy of
Sciences, Grant No. XDB23000000.

\section*{Data Availability}
This theoretical study did not generate any new data.


\bibliographystyle{mnras}
\bibliography{ref}



\appendix

\bsp	
\label{lastpage}
\end{document}